\documentclass[journal, final, 10pt, twocolumn]{IEEEtran}
\usepackage{layout}
\usepackage[letterpaper, left=.92in, right=.92in, bottom=.7in, top=0.70in]{geometry}

\def\bb0{{\mathbb{0}}}

\def\bb{{\boldsymbol{b}}}

\def\b0{{\boldsymbol{0}}}

\def\s{{\mathrm{s}}}
\def\x{{\mathrm{x}}}
\def\y{{\mathrm{y}}}

\def\b{{\mathrm{b}}}

\def\r0{{\mathbf{0}}}

\def\cE{\mathcal{E}}

\def\cH{\mathcal{H}}

\def\cQ{\mathcal{Q}}

\def\bsf0{{\bm{\mathsf{0}}}}

\def\c{{\rm c}} 

\def\N0{{N_{\mathrm{0}}}}

\newcommand{\Ncgauss}{\mathcal{N}_{\mathbb{C}}}

\def\fc{f_{\mathrm{c}}}

\def\bsf{{\boldsymbol{s}_\mathrm{f}}}

\newcommand{\be}{\begin{equation}}
\newcommand{\ee}{\end{equation}}
\newcommand{\bal}{\begin{align}}
\newcommand{\eal}{\end{align}}

\def\Exp{{\mathbb{E}}}
\def\C{{\mathsf{C}}}

\def\Ecal {\mathcal{E}}

\def\SNR    {{\mathsf{SNR}}}
\def\SDNR    {{\mathsf{SNDR}}}

\def\fc {f_{\mathrm{c}}}

\usepackage[svgnames]{xcolor}
\usepackage{graphicx}
\usepackage{amsmath}
\usepackage{subfig}
\usepackage{mathtools}
\usepackage{epsfig}
\usepackage{float}
\usepackage{lipsum}
\usepackage{stfloats}
\usepackage{array}
\usepackage{amssymb}
\usepackage{cite}
\usepackage{url}
\usepackage{amsthm}
\usepackage{algorithm}
\usepackage{algorithmic}
\usepackage{multirow}
\usepackage{flushend}
\usepackage{upgreek}
\usepackage{diagbox}
\usepackage{dsfont}
\usepackage{xifthen}
\usepackage{siunitx}
\usepackage{tabularx}
\usepackage{colortbl}
\definecolor{NYU_purple}{rgb}{0.54,0.35,0.71}

\usepackage{tikz}
\usetikzlibrary{chains,arrows,calc,positioning}
\usetikzlibrary{shapes.multipart}
\usetikzlibrary{decorations.pathreplacing}
\usetikzlibrary{calc}
\usetikzlibrary{shapes.geometric}

\newtoggle{arxiv}
\toggletrue{arxiv}

\normalsize
\allowdisplaybreaks

\newcommand{\GD}[1][]{3
	\ifthenelse{\isempty{#1}}%
	{\text{G}_{i,j}^k}
	{{(\text{G}_{i,j}^k)}^{#1}}
}

\theoremstyle{definition}

\usepackage{layout}
\usepackage{bm}
\usepackage{comment}
\usepackage{color,soul}
\usepackage{makecell}
\usepackage{amssymb}

\def\*#1{\mathbf{#1}}
\def\w*#1{\widehat{#1}}

\def\s*#1{\mathsf{#1}}
\def\S*#1{\bm{\mathsf{#1}}}

\def\c*#1{\mathcal{#1}}
\def\C*#1{\bm{\mathcal{#1}}}
\def\T*#1{\text{#1}}

\newcommand{\subsf}{\sf \scriptscriptstyle}

\begin{document}

\title{Spectral vs Energy Efficiency in 6G: \\ Impact of the Receiver Front-End
\thanks{A. Lozano is with Universitat Pompeu Fabra, 08018 Barcelona (e-mail: {\tt angel.lozano@upf.edu}). His work is supported by the Fractus-UPF Chair on Tech Transfer and 6G, by the ICREA Academia program, and by the Spanish research agency through project PID2021-123999OB-I00.
S. Rangan is with New York University,
Brooklyn, BY (e-mail: {\tt srangan@nyu.edu}).
His work is supported in part by NSF grants
1952180, 2133662, 2236097, 2148293, and 1925079,
along with the industrial affilates of NYU Wireless.}
}
\author{\IEEEauthorblockN{Angel~Lozano}, {\it Fellow,~IEEE},
\and
\IEEEauthorblockN{Sundeep~Rangan},
{\it Fellow,~IEEE}
}

\maketitle

\begin{abstract}
This article puts the spotlight on the receiver front-end (RFE), an integral part of any wireless device that information theory typically idealizes into a mere addition of noise.
While this idealization was sound in the past, as operating frequencies, bandwidths, and antenna counts rise, a soaring amount of power is required for the RFE to behave accordingly. Containing this surge in power expenditure exposes a harsher behavior on the part of the RFE (more noise, nonlinearities, and coarse quantization), 
setting up a tradeoff between the spectral efficiency under such nonidealities and the efficiency in the use of energy by the RFE.
With the urge for radically better power consumptions and energy efficiencies in 6G, 
this emerges as an issue on which information theory can cast light at a fundamental level.
More broadly, this article advocates the interest of having information theory embrace the device power consumption in its analyses. In turn, this calls for new models and abstractions such as the ones herein put together for the RFE, and for a more holistic perspective.

%

\end{abstract}

\section{Motivation}
\label{sec:motivation}

Energy efficiency was an important driver in the design of 1G and 2G wireless standards where, for instance, more efficient amplifiers were enabled by the adoption of (respectively analog and digital) signaling formats tolerant of nonlinear amplification.
By the time 3G came to be, however, the perception of spectrum scarcity in the low frequency bands had brought about a shift in priorities, and
energy efficiency has since taken a back seat. 
Indeed, a chief thrust in 3G, 4G, and 5G, was to improve the spectral efficiency. 

In the post-5G era, the pendulum is swinging back---not so much because the demand for spectral efficiency is relenting, but because
energy efficiency is becoming an imperative. The ICT sector is anticipated to devour a staggering $20\%$ of the global electricity production by 2030 and, homing in on wireless networks, the radio access is the most energy-hungry portion \cite{Wesemann2023}.
Meanwhile, at the device end, autonomy and battery life are expected to acquire paramount importance.
This mounting pressure on the use of energy
brings renewed interest in the tradeoff between spectral and energy efficiency, for years skewed all the way towards the former.

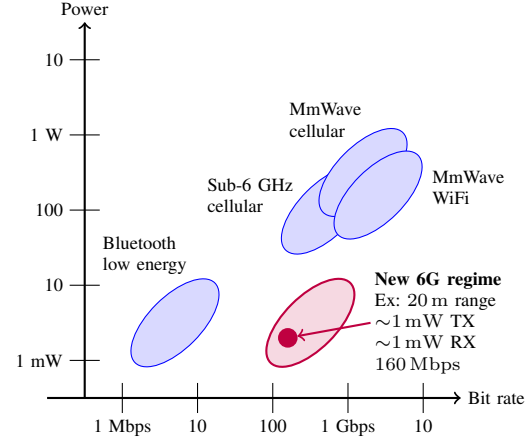
\begin{figure}
\scriptsize
\centering
\begin{tikzpicture}

\draw[thick,->] (-1,-0.5) -- (4.5,-0.5) node [right]
{Bit rate};
\draw[thick,->] (-0.5,-0.5) -- (-0.5,4.5) node [above] {Power};

\foreach \x/\txt in {0/{1 Mbps}, 1/10,2/100,3/{1 Gbps},4/10} {
    \draw [-] (\x,-0.3) -- (\x,-0.7)
    node [below] {\txt};
};
\foreach \y/\txt in {0/{1 mW}, 1/10,2/100,3/{1 W},4/10} {
    \draw [-] (-0.3,\y) -- (-0.7,\y)
    node [left] {\txt};
};


\foreach \x/\y in {0/0, 2/1.5, 2.5/2, 2.7/1.7}  {
    \pgfmathsetmacro{\xa}{\x + 0.7};
    \pgfmathsetmacro{\ya}{\y + 0.5};
    
    \node[ellipse,blue,draw,rotate=45,
        fill=blue!15,
        minimum width=1.5cm,minimum height=0.7cm,
        fill opacity=0.3] at (\xa,\ya) {}; 
}

\node[ellipse,purple,thick,draw,rotate=45,
        fill=purple!15,
        minimum width=1.5cm,minimum height=0.7cm] 
        at (2.5,0.5) {};
\node[ellipse,purple,thick,draw,
        fill=purple] 
        at (2.2,0.3) (pt) {};

\node[align=left] at (4.2,0.5) {\textbf{New 6G regime} \\
Ex: \SI{20}{m} range \\
    $\sim$\SI{1}{mW} TX\\
    $\sim$\SI{1}{mW} RX\\
    \SI{160}{Mbps}};
\draw[->,purple,thick] (3.3,0.5) -- (pt);

\node[align=left] at (0.3,1.4) {Bluetooth \\ low energy};
\node[align=left] at (1.7,2.2) {Sub-6 GHz\\cellular};
\node[align=left] at (2.7,3.2) {MmWave\\cellular};
\node[align=left] at (4.6,2.3) {MmWave\\WiFi};

\end{tikzpicture}
\caption{Power and rate regions
of operation of existing wireless technologies, 
alongside the new low power, high bit rate regime of potential interest 
for 6G.  An example in this regime is pinpointed
based on the 
models in this paper combined with 
current RF device power vs.\ performance characteristics (see Appendix~\ref{sec:usecases} for details). }
\label{fig:regimes}
\end{figure}

At the same time that they regain importance, energy efficiency assessments need to become more holistic.
Classically, the only power component that information theory has concerned itself with is the transmit power,
but with the progression towards ever higher carrier frequencies, much broader bandwidths, and multiplied antenna counts, the power consumed by the circuitry is bound to become comparable to the transmit power at the infrastructure end \cite[Fig. 1]{Wesemann2023} and might outright dwarf it
at many devices.




By far the biggest contributor to the circuit power consumption is certain to be the receiver front-end (RFE),
whose consumption is swelling already in 5G.
For 6G, carrier frequencies could reach $300$ GHz, with multi-gigahertz bandwidths and on the order of $64$ antennas at mobile units \cite{Wesemann2023}, posing a major challenge.
%
%
%
Even for lower frequencies, bandwidths, and antenna numbers, the RFE consumption becomes an issue if high autonomies and/or miniature batteries are desired.
As illustrated in Fig.~\ref{fig:regimes},
in contrast with existing paradigms of high bit rates 
with a high transmit power
or low bit rates 
with a low transmit power, many 6G devices may be in a new class of
their own: high bit rates over short distances
with a low transmit power.
In such devices, the RFE power consumption might overshadow the transmit power \cite{mezghani2011power},
and lowering the former would enable
a powerful regimes of use cases for applications such as 
smart wearables, untethered cameras, virtual reality goggles, connectivity modules, or compact short-range access points devoid of cooling systems.

\iftoggle{arxiv}{Such devices may indeed be possible. As detailed in Appendix~\ref{sec:usecases}, 
leveraging rigorous information theoretic models with 
current device performance vs.\ power behaviors,
it may be possible, as an example, to build a transceiver
achieving \SI{160}{Mbps} with as little as 
\SI{1}{mW}
of transmit power, \SI{1}{mW} of power consumption
at the receiver, and \SI{20}{m} of range in a typical
indoor environment.    
Such performance numbers would be dramatically
outside the power-bit rate characteristics
of existing wireless systems.}{}

Crucially, any effort to tame the power expended by an RFE pushes it away from ideality, aggravating its imperfections: higher noise floor, nonlinear behavior, and coarser quantization \cite{skrimponis2020towards}.
Characterizing the impact of the RFE on the fundamental performance of a communication system requires models that abstract these essential aspects of noisiness, nonlinearity, and quantization coarseness in a manner that is useful from an information-theoretic vantage, as well as a physically-based model for the concomitant power consumption.
Coarse quantization has been dealt with extensively already, and important results have emerged \cite{singh2009limits}.
This article expands the modeling scope to also subsume the other aspects.

Armed with a model for the RFE, both the spectral efficiency that it enables (in bits/s/Hz) and its energy efficiency (in bits/s/Watt or bits/Joule) can be characterized, with the latter also expressible through its reciprocal, the energy per bit (in Joules/bit).
The spectral efficiency and the RFE energy per bit inform, respectively, of the utilization of the bandwidth and of the energy stored by the device's battery.
These two performance measures are informative on their own, yet
they are even more revealing when pitted against each other.
This amounts to expressing the spectral efficiency, not as a function of the energy
expended by the RFE per unit of time, but per unit of information.
Indeed, pushing the RFE towards ideality increases its power consumption, but that may be worthwhile if sufficiently more bits can then be sent through.
Conversely, relaxing the RFE specifications lowers its consumption, but that need not be fruitful from an energy per bit standpoint, depending on how much the bit rate drops.
A spectral-vs-energy efficiency assessment is the appropriate framework to make these determinations, examine the interplay of the RFE key knobs, and glean design guidelines that explicitly account for the energy cost of concealing the RFE's imperfections.

These RFE imperfections, which motivate the analysis in the first place, also complicate it, as they give rise to settings that deviate from the familiar linear channel with additive white Gaussian noise (AWGN). These broader settings are described, relevant works are surveyed, and open issues are identified.

%
%

Altogether, this article contemplates the impact of the RFE in potential 6G designs operating at higher
carrier frequencies, much wider bandwidths, increased antenna counts, and/or with far lower power consumption limits and energy budgets.
%
%
%
Looking beyond, the formulation could be augmented with other expanding contributions to the power consumption (say the channel decoder \cite{korb2010ldpc}), for an ever more comprehensive assessment of the cost of operating a receiver. This could then be conceivably blended with its transmitter counterpart.
Indeed,
in terms of efficiency what matters is the total expended energy \cite{murdock2013consumption,kanhere2022power}; holistic assessments align with 3GPP's energy consumption and efficiency metrics, which account for the total energy irrespective of where it is consumed. 

\iftoggle{arxiv}{}{For proofs of some of the results and further considerations, readers are referred to \cite{lozano2023spectral-arxiv}.  }

\section{RFE Model} \label{sec:model}

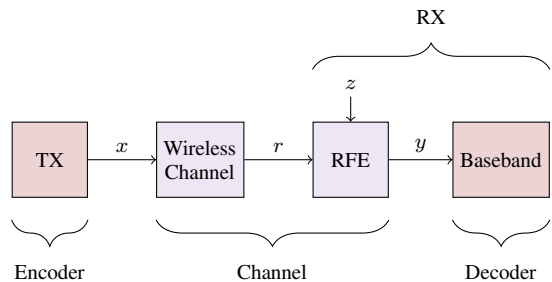
\begin{figure}
\centering
\begin{tikzpicture}[scale=1.0,every text node part/.style={align=center}, every node/.append style={transform shape}]
    \footnotesize
    
    \node [draw,minimum width=1cm, minimum height=1cm, fill=Brown!20, node distance=1cm]    (TX) {TX};
    \node [draw,,minimum width=1cm, minimum height=1cm, 
        fill=NYU_purple!15, node distance=1cm, right of=TX,
        xshift=1cm]    (Chan) {Wireless\\ Channel};
    \node [draw,,minimum width=1cm, minimum height=1cm, 
        fill=NYU_purple!15, node distance=1cm, right of=Chan,
        xshift=1cm]    (RFFE) {RFE};
    \node [draw,,minimum width=1cm, minimum height=1cm, 
        fill=Brown!20, node distance=1cm, right of=RFFE,
        xshift=1cm]    (RX) {Baseband};

    \node [above of=RFFE] (noise) {$z$};

    \draw [->] (TX) -- (Chan) node[midway,above] {$x$};
    \draw [->] (Chan) -- (RFFE) node[midway,above] {$r$};
    \draw [->] (RFFE) -- (RX) node[midway,above] {$y$};
    \draw [->] (noise) -- (RFFE);

    \draw [decorate,
        decoration={brace,amplitude=10pt,raise=1.2cm}] 
        (RFFE.west) -- node[above,yshift=1.7cm] {RX} (RX.east) ;
    \draw [decorate,
        decoration={brace,amplitude=10pt,raise=0.8cm,mirror}] 
        (Chan.west) -- node[below,yshift=-1.3cm] {Channel} (RFFE.east);
    \draw [decorate,
        decoration={brace,amplitude=10pt,raise=0.8cm,mirror}] 
        (RX.west) -- node[below,yshift=-1.3cm] {Decoder} (RX.east);
    \draw [decorate,
        decoration={brace,amplitude=10pt,raise=0.8cm,mirror}] 
        (TX.west) -- node[below,yshift=-1.3cm] {Encoder} (TX.east);

\end{tikzpicture}

\caption{Transmission chain with transmitter, channel, and receiver. From the information-theoretic vantage of encoding and decoding, the RFE can be subsumed into the channel.}
\label{fig:model}

\end{figure}

For starters, let us consider the setting in Fig.~\ref{fig:model},
with a single-antenna receiver and bandwidth $B$.
The discrete-time complex baseband transmit signal is $x$
while the noiseless received signal is $r=h x$
%
for a given complex channel gain, $h$. The energies per symbol are
$\Ecal_x = \Exp[|x|^2]$ and $\Ecal_r = \Exp[|r|^2] =  \Exp [|h|^2] \cE_x$. 

\subsection{Linear RFE Model}

A wireless receiver consists of two stages: the RFE, which effects the downconversion, filtering, and digitalization, and the baseband processor, which demodulates and decodes. 
The standard model for the RFE is 
\begin{equation} \label{eq:rxawgn}
    y = r + z, 
\end{equation}
where $z$ is complex Gaussian noise with 
variance $kTF$; here, $kT$ is the minimum theoretical value ($-174$ dBm/Hz
at room temperature) while $F>1$ is the noise figure quantifying the increase in noise due to RFE nonideality. With that, the signal-to-noise ratio is $\SNR = \SNR_{\sf ideal} / F$ given
\begin{equation} \label{eq:snrf}
    \SNR_{\sf ideal} = \frac{\cE_r}{kT} .
\end{equation}
This AWGN setting has spawned much of the wisdom on the fundamental limits of reliable communication. In particular, 
the highest achievable bit rate is 
\be
R = B \log_2(1+\SNR),
\ee
attained when $x$ is complex Gaussian.
The corresponding spectral efficiency is $C=R/B$.

Having the RFE approach the behavior in \eqref{eq:rxawgn} is however very costly in terms of power consumption
and, as advanced, as one attempts to lower that cost, the behavior becomes less benign. This less benign functioning can be captured by the broader model $y = \Phi(r, z)$,
where $\Phi(\cdot)$ is a memoryless, generally nonlinear function.





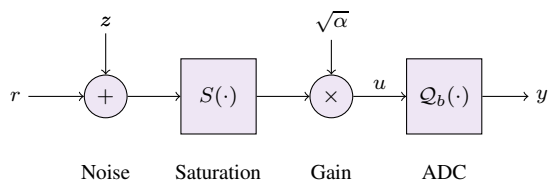
\begin{figure}
\centering
\begin{tikzpicture}[scale=1.0,every text node part/.style={align=center}, every node/.append style={transform shape}]
    \footnotesize
    
    \node [draw,circle,fill=NYU_purple!15]  (dplus)  {$+$};
    \node [left of=dplus,xshift=-0.2cm] (r) {$r$};            
    \node [above of=dplus] (noise) {$z$};  
    \node [draw,minimum width=1cm, minimum height=1cm,  
        fill=NYU_purple!15, right of=dplus,xshift=0.5cm]   
        (sat) {$S(\cdot)$};  
    \node [draw,circle,fill=NYU_purple!15,right of=sat,
        xshift=0.5cm] (gainq) {$\times$};          
    \node [draw,minimum width=1cm, minimum height=1cm,  
        fill=NYU_purple!15,right of=gainq,
        xshift=0.5cm]    (quant) {$\cQ_b(\cdot)$};    
    \node [right of=quant,xshift=0.3cm] (y) {$y$};

    \node [above of=dplus] (noise) {$z$};
    \node [above of=gainq] (alpha) {$\sqrt{\alpha}$};

    \draw [->] (r) -- (dplus);
    \draw [->] (dplus) -- (sat);
    \draw [->] (sat) -- (gainq);
    \draw [->] (gainq) --  node [midway,above] {$u$} (quant);
    \draw [->] (quant) -- (y);
    \draw [->] (noise) -- (dplus);
    \draw [->] (alpha) -- (gainq);

    \node [below of=sat, yshift=-0.0cm]  {Saturation};
    \node [below of=quant, yshift=-0.0cm]  {ADC};
    \node [below of=gainq, yshift=-0cm]  {Gain};
    \node [below of=dplus, yshift=0cm]  {Noise};
    
\end{tikzpicture}

\caption{RFE modeled as a cascade of four  operations: additive noise, saturation, gain, and quantization.}
\label{fig:rffe}

\end{figure}

\subsection{RFE Operations}

The specific form for $\Phi(\cdot)$ propounded here is, as illustrated in Fig.~\ref{fig:rffe},
\begin{align} 
    \Phi(r,z) =  \cQ_b \! \left( \sqrt{\alpha}  S(r + z) \right) ,  \label{eq:genrx}
\end{align}
which expresses a cascade of components, each a critical feature of the RFE as
detailed next.


\begin{itemize}

\item \emph{Additive noise}:  As in the linear model, $z$ is the noise, with
variance $kTF$.

\item \emph{Saturation.}  The transformation $S(\cdot)$ brings into the
    model the finite linear range 
    via
\begin{equation} \label{eq:satdef}
    S(\xi) = \sqrt{\cE_{\sf max}} \, \phi \! \left(\frac{|\xi|}{\sqrt{\cE_{\sf max}}}\right)
    \frac{\xi}{|\xi|},   
    \end{equation}
    where $\phi(\cdot) \in (0,1]$ is a function satisfying $\phi(0) = 0$ and $\phi'(0) = 1$.
For the sake of analysis, two concrete such functions can be $\phi(\cdot) = \tanh(\cdot)$, which is smooth, and $\phi(\cdot) = \min(\cdot,1)$, an outright clipping.
The transform behaves linearly for small inputs, but it saturates at
a signal power of $P_{\sf max} = \cE_{\sf max} B$.


   
\item \emph{Quantization.}  The function $\cQ_b(\cdot)$
    models the analog-to-digital converter (ADC), with
    $b$ bits per each of the in-phase and quadrature dimensions.
    
\item \emph{Gain control.}  
Critical to proper operation of the ADC is a scaling of its input
to prevent overflow or underflow.
This is represented by $\alpha$. 

\end{itemize}




Altogether, the RFE is characterized by the knobs $(F, P_{\rm max}, b)$.
In addition, the performance is influenced by the strategy for setting the gain control.


\subsection{Power Consumption}


There is considerable range in the RFE power consumption for a set of 
performance attributes, with deviations due to the process technology,
form factor, and other considerations.
For the purpose here, what is sought is not a model to 
make precise consumption predictions, but to---based on the scaling laws of circuits and devices---abstract the dependencies on the RFE knobs and on the system parameters, chiefly frequency and bandwidth.

The noise figure is determined primarily by the low-noise amplifier 
 (LNA), at whose input the signal is weakest. The most widely used scaling for the LNA power consumption is with $\fc/(F-1)$ where $\fc$ is the carrier frequency \cite{song2008simple};
this is the scaling adopted here, even if \cite{belostotski2021figures}
suggests that, at very low powers, the scaling may be more aggressive, namely with $\fc^2/(F-1)^3$. As of the saturation, it occurs in both the LNA and mixer, and the incurred power is reasonably modeled as proportional to $P_{\sf max}$ \cite{brederlow2001mixed,ramzan2012figure,chen2006design,verma202136}.
With no single accepted rule for how the noise figure and the saturation power consumptions should be combined
\cite{ramzan2012figure,verma202136},
this text espouses their addition; this conservative choice is sure to be valid for moderate perturbations around nominal values of $F$ and $P_{\sf max}$.

Then, the two ADCs required to process a complex signal consume a power that scales with $B \kappa^b$ where $\kappa \approx 4$ for signal-to-quantization ratios in excess of $40$--$50$~dB while $\kappa \approx 2$ for the lower ratios at which wireless systems operate \cite{murmann2016adc}; $\kappa = 2$ is thus considered in the sequel.
Although, beyond a few hundred megahertz, the ADC consumption would become quadratic in the bandwidth, the linear scaling can always be retained through parallelization and ADC pipelining (multiple low-resolution stages cascaded to obtain a higher resolution). Only if it were truly necessary to quantize a single band exceeding hundreds of megahertz would a quadratic behavior be experienced.

The gain control, finally, is typically performed at baseband and consumes negligible power relative to the rest.  
%
All in all, the proposed model for the RFE power consumption is
\begin{equation} \label{eq:Pcons}
    P_{\sf \scriptscriptstyle RFE} = \underbrace{ \gamma_{\sf \scriptscriptstyle NF} \frac{\fc}{F-1} }_{\text{noise figure}} + \underbrace{ \gamma_{\sf max}  P_{\sf max} }_{\text{saturation}} + \underbrace{ \gamma_{\sf \scriptscriptstyle ADC} B \, 2^b }_{\text{ADC}}
\end{equation}
with indicative values for the figures of merit $\gamma_{\sf \scriptscriptstyle ADC}$, $\gamma_{\sf \scriptscriptstyle NF}$, and $\gamma_{\sf max}$, presented in Table~\ref{tab:powparam} . In particular, $\gamma_{\sf \scriptscriptstyle ADC}$ improves by about $1.5$ dB per year, with a fundamental limit imposed by physics anticipated at $\gamma_{\sf \scriptscriptstyle ADC} \approx 0.1$ fJ/qt \cite{murmann2016adc}.
\iftoggle{arxiv}{See Appendix~\ref{sec:PFEjustification} }{See \cite{lozano2023spectral-arxiv} }
for more details on the figures of merit.



Interestingly, the power consumption term induced by each knob involves a distinct system parameter:
\begin{itemize}
\item The power that must be burned to attain a certain noise figure is determined by the carrier frequency, but neither the bandwidth nor the received signal strength. And, rewriting it as
\begin{align}
    \frac{\gamma_{\sf \scriptscriptstyle NF} \fc}{F-1} = \frac{\gamma_{\sf \scriptscriptstyle NF} \fc}{\frac{\SNR_{\sf ideal}}{\SNR}-1} ,
\end{align}
it cleanly connects the consumed power, the carrier frequency, and the 
SNR degradation.
\item The power that needs to be expended to stretch the range of unsaturated signals depends on the received signal strength, and hence it relates to $\SNR_{\sf ideal}$.
\item The power consumed to operate at a certain ADC resolution is contingent on the bandwidth, but not on the carrier frequency or the received signal strength.
\end{itemize}

\begin{table}[t]
    \centering
    \begin{tabularx}{\linewidth}{|l|l|>{\hsize=0.5\hsize\raggedright}X|>{\hsize=1.5\hsize\raggedright}X|} 

    \hline
    \rowcolor{NYU_purple!15} 
    Figure of Merit & Units & Value &  Remark \tabularnewline \hline  
    $\gamma_{\sf \scriptscriptstyle ADC}$ & \si{fJ/qt} & 165 &  Based on ADC  data \cite{murmann2016adc} 
    \tabularnewline \hline  

    $\gamma_{\sf \scriptscriptstyle NF}$ & \si{fJ} & 140 &  Based on LNA data \cite{belostotski2021figures} \tabularnewline \hline  

    $\gamma_{\sf max}$ & - & $5000$ &  Based on mixer data \cite{wang200951,skrimponis2020towards,guo2014wideband} \tabularnewline \hline  
    \end{tabularx}
    \caption{Figures of merit based on recent
    device and circuit surveys.
    }
    \label{tab:powparam}
\end{table}


\section{Spectral and Energy Efficiency}

When a nonlinear transformation sits between encoder and decoder, the noise is compounded by distortion and quantifying the information-theoretic performance limits, as well as strategies that can approach them, becomes notoriously difficult.
In such a context, a very useful result can be formulated on the basis of
the signal-to-noise-and-distortion ratio
\be
\SDNR = \frac{\rho^2_{xy}}{1-\rho^2_{xy}} 
\label{SNDR}
\ee
where noise and distortion are blended via
the squared correlation coefficient
\begin{equation} \label{eq:rhoxy}
    \rho^2_{xy} = \frac{\left|\Exp [ x^* y ] \right|^2}{\cE_x \, \cE_y} .
\end{equation}
Based on the $\SDNR$, the spectral efficiency with complex Gaussian signaling
satisfies \cite{6109374,dutta2020capacity} 
\be
    C \geq \log_2(1+\SDNR).
\label{Rome}
\ee
This bound can be
%
%
obtained from the Bussgang-Rowe decomposition, whereby the output of a nonlinear transformation is split into a scaled version of the input plus uncorrelated distortion (in general neither Gaussian nor independent of the input); the bound arises by then replacing that distortion with Gaussian noise of the same variance.

When a unitary transformation is applied between the RFE and the decoder, for instance the time-frequency transformation in multicarrier signaling, the distortion is thrown around and asymptotically (in the dimensions of the unitary transformation, say the number of subcarriers) it is rendered Gaussian; then, the bound gives the actual achievable spectral efficiency \cite{dutta2020capacity}.
Without a post-RFE transformation, the distortion remains signal-dependent and non-Gaussian; these attributes could conceivably be taken advantage of by the decoder, rendering the lower bound somewhat conservative.
Customarily though, the decoder treats the distortion as additional Gaussian noise, whereby its effect (on Gaussian signals) is indeed that of additional Gaussian noise \cite{lapidoth1996nearest,720535}. 
Motivated by this argument, the lower bound in (\ref{Rome}) is regarded as the achievable spectral efficiency in the sequel.

As of the RFE's energy efficiency, it equals $R/P_{\sf \scriptscriptstyle RFE}$. At the infrastructure end, this is often measured in Mb/kWh, reflecting that electricity is billed in kWh, while at mobile devices b/J is a more fitting unit. More germane to information theory is actually its reciprocal, the RFE energy per bit
\be
\cE^{\sf \scriptscriptstyle RFE}_{\rm b} = \frac{P_{\sf \scriptscriptstyle RFE}}{R} = \frac{1}{C}
\left(
\frac{\gamma_{\sf \scriptscriptstyle NF}}{F-1} \frac{\fc}{B} + \gamma_{\sf max}  \cE_{\sf max} + \gamma_{\sf \scriptscriptstyle ADC}  2^b
\right)
\label{Aitana}
\ee
in J/b; this is the measure adopted henceforth. 
Note that we have used the fact that 
$P_{\subsf max} = \cE_{\subsf max}B$.
Importantly, $\cE^{\sf \scriptscriptstyle RFE}_{\rm b}$, which relates to the RFE power consumption, should not be confused with the transmit and received energy per bit, both associated with the radiated power \cite[Sec. 4.2]{heath2018foundations}.

An important implication of \eqref{Aitana}
is that the dependence of the 
RFE energy per bit 
on the bandwidth and carrier frequency is
only through the ratio, $B/f_c$.

\subsection{A Closer Look at the $\SDNR$}

\begin{figure}
    \centering
    \includegraphics[width=\linewidth]{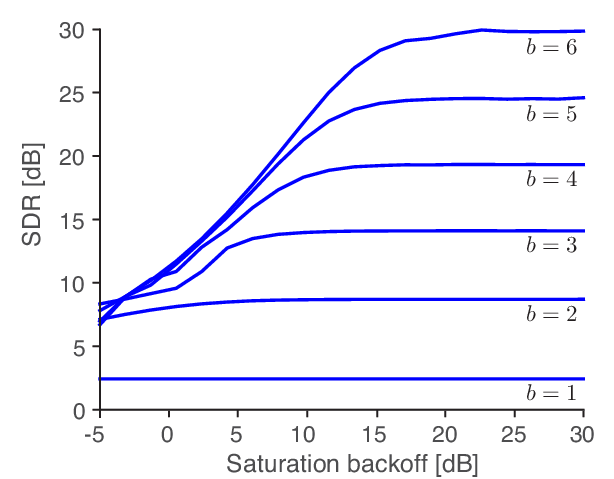}
    \caption{SDR as a function of the ADC resolution and the saturation backoff.}
    \label{fig:snrmaxqsat}
\end{figure}

Without loss of generality, the ADC can be designed for a unit-energy input, whereby the required gain control is
\begin{equation} \label{eq:alphaq}
\alpha = \frac{1}{\Exp[|S(r+z)|^2]} .
\end{equation}
With this gain, and given how $x$, $r$, and $z$ are related through $\SNR_{\sf ideal}$, (\ref{SNDR}) can be manipulated
\iftoggle{arxiv}{(See Appendix~\ref{sec:sndrsat_deriv})}{
(See \cite{lozano2023spectral-arxiv})}
into
\begin{equation}  \label{eq:sndrsat}
    \SDNR = \frac{\SNR_{\sf ideal} \, \rho^2(b,\nu)}{F + \SNR_{\sf ideal} \, (1-\rho^2(b,\nu))}, 
\end{equation}
where
\begin{align} \label{eq:rhobnu}
    \rho^2(b,\nu)    = \frac{|\Exp [ (r+z)^* y ] |^2}{(\cE_r+N_0) \, \cE_y} 
\end{align}
implicitly depends on the function $\phi(\cdot)$ in \eqref{eq:satdef} and must generally be computed numerically. Interestingly, $\rho(b,\nu)$ depends
only the ADC resolution and on the saturation backoff, the latter given by
\begin{equation} \label{eq:nudef}
    \nu = \frac{\Ecal_{\sf max}}{\Ecal_r + N_0} .
\end{equation}
No matter how high $\SNR_{\sf ideal}$, the $\SDNR$ is curbed, namely
\be
\SDNR \leq \frac{\rho^2(b,\nu)}{1-\rho^2(b,\nu)} 
\label{curbed}
\ee
whose right-hand side is the noiseless signal-to-distortion ratio (SDR).
Fig.~\ref{fig:snrmaxqsat} shows this SDR for $\phi(\cdot) = \tanh(\cdot)$, illustrating the effects of the resolution and the saturation backoff.

\subsection{Only Noise and Quantization}

When the backoff is large, the input is always below the saturation level
and the distortion is caused only by the ADC. 
Then, the gain control becomes $\alpha = 1/(\Ecal_r + N_0)$
and
(\ref{eq:sndrsat}) continues to hold, only with $\rho(\cdot)$ a function of solely the resolution, namely
\begin{align}
    \rho^2(b)    = \frac{|\Exp \! \left[ u^* \cQ_b(u) \right] |^2}{\Exp \! \left[|\cQ_b(u)|^2 \right]}
    \label{Alamo}
\end{align}
where $u \sim \Ncgauss(0,1)$ represents the gain-controlled signal being fed to the ADC.
Since $\rho(b) \in [0,1]$,
\be
0 \leq
\SDNR \leq \frac{\SNR_{\sf ideal}}{F}
\ee
as one would expect, with the highest value corresponding to infinite resolution.
%
%
At low $\SNR_{\sf ideal}$,
\begin{equation}
    \SDNR \approx \SNR_{\sf ideal} \frac{\rho^2(b)}{F},
\end{equation}
indicating that the noise figure and the ADC affect the $\SDNR$ in a multiplicative fashion.
%
At high $\SNR_{\sf ideal}$, in turn, the $\SDNR$ is bounded  by the SDR as per (\ref{curbed}).


\subsection{Optimum Quantization}

For a complex Gaussian signal, the quantization distortion is minimized, in the mean-square sense, by a vector quantizer operating over asymptotically long blocks
and having itself a complex Gaussian codebook \cite{gersho2012vector}. Then, the distortion is itself Gaussian and independent of the quantized signal, and 
\be
\cQ(u) = \sqrt{1-2^{-2b}} \left( \sqrt{1-2^{-2b}} u + d \right)
\label{SCopa}
\ee
with $d \sim \Ncgauss \! \left(0,  2^{-2b}   \right) $;
this relationship evinces a loss in signal energy and the appearance of the quantization distortion $d$, which in this case does amount to additional Gaussian noise.
Note that, with vector quantization over asymptotically long blocks, $b$ embodies the average number of bits per symbol, which need not be integer. The number of bits $M b$ representing a block of $M$ symbols is to be an integer, but, for $M \to \infty$, that allows $b$ to take any rational value.
From (\ref{Alamo}) and (\ref{SCopa}), 
\be
    \rho^2(b) = 1 - 2^{-2b}  ,
    \label{anton}
\ee
which plugged into (\ref{eq:sndrsat}) gives the $\SDNR$.
Moreover, with the distortion introduced by the optimum vector quantizer being complex Gaussian and independent of the signal, the bound in (\ref{Rome}) is then the actual spectral efficiency achievable with complex Gaussian signaling, even without a post-RFE unitary transformation.
However, the situation is not akin to an AWGN setting because, for a given resolution, the $\SDNR$ is curbed and thus caution must be exercised when applying known results, in particular notions such as the degrees-of-freedom that are inherently asymptotic.

\begin{figure}
    \centering
    \includegraphics[width=\linewidth]{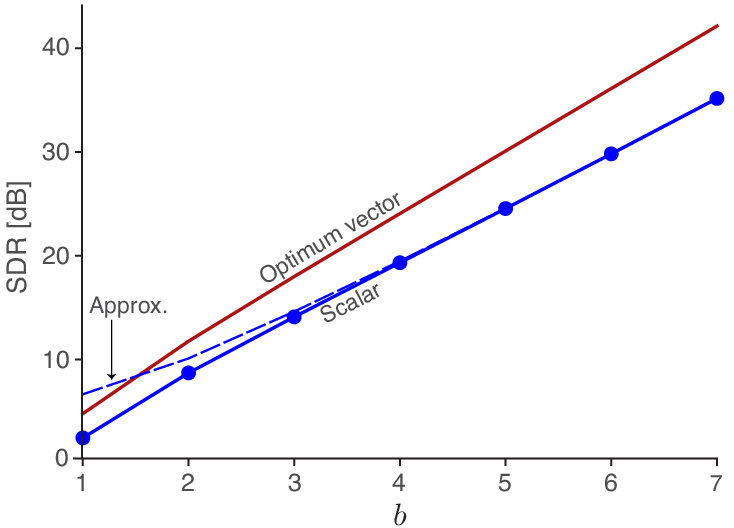}
    \caption{Noiseless SDR as a function of the ADC resolution: optimum vector quantization vs scalar uniform quantization and its asymptotic approximation in (\ref{Patri}).}
    \label{fig:snrmaxquant}
\end{figure}

Interestingly, the optimum vector quantizer turns out to be a remarkably faithful representation of the scalar uniform quantizers that are preferred from an implementation standpoint, with $b$ integer-valued. 
Although, in general, the ensuing $\rho^2(b)$ has to be determined numerically,
for growing $b$ \cite{hui2001asymptotic}
\be
\rho^2(b) \approx 1- c \, b \, 2^{-2b}
\label{dominant}
\ee
for some constant $c$. 
Consequently,
\begin{align}
\SDNR \approx \frac{(1 - c \, b \, 2^{-2b}) \, \SNR_{\sf ideal}}{ F + c \, b \, 2^{-2b} \, \SNR_{\sf ideal} } .
\label{Patri}
\end{align}
Presented in Fig.~\ref{fig:snrmaxquant} is a comparison between the noiseless SDR with optimum vector quantization and with scalar uniform quantization;
also shown is how (\ref{Patri}) matches the latter for $b \geq 3$, requiring only the calibration of $c$.
When the backoff is sufficient and the distortion is introduced solely by the ADC, information-theoretic analyses are thereby enabled with only a correction for $b \leq 2$.


%

\section{Design Guidelines: A Case Study}

\label{sec:powopt}

\begin{figure}
    \centering
    \includegraphics[width=\linewidth]{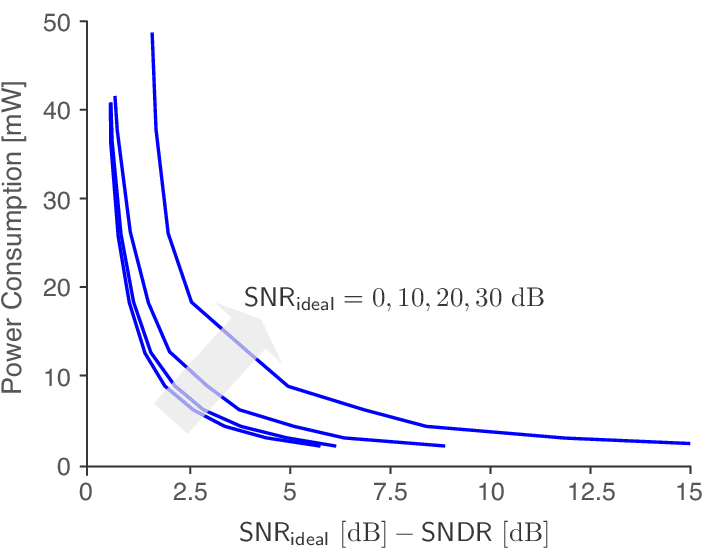}
    \caption{Minimum power consumption over all RFE configurations as a function of the dB-difference between $\SNR_{\sf ideal}$ and $\SDNR$ for $\fc=28$ GHz, $B=400$ MHz, and scalar uniform quantization.}
    \label{fig:snr_loss}
\end{figure}

To make the analysis concrete, consider
an exemplary mmWave system with $f_c=$\,\SI{28}{GHz}
and $B=$\,\SI{400}{MHz}, corresponding to $B/f_c=1/70$.
The behavior of the $\SDNR$ in this case is 
presented in Fig.~\ref{fig:snr_loss}, which shows, for every value (referenced to $\SNR_{\sf ideal}$), the minimum power consumption over all combinations of $(F,P_{\sf max},b)$. 
The picture vividly illustrates the soaring power required to bring the $\SDNR$ ever closer to $\SNR_{\sf ideal}$, even in the best possible configuration of the RFE. Also noteworthy is how the power consumption shifts up with $\SNR_{\sf ideal}$, as better---hence more costly in terms of power---noise figures, saturation levels, and resolutions can be capitalized on.
This is a manifestation of the intricate relationship between
operating point and RFE knobs that is explored in this section, as an illustration of how the RFE model and the performance measures derived from it can be put to use.


\begin{figure*}
    \centering
    \includegraphics[width=16cm]{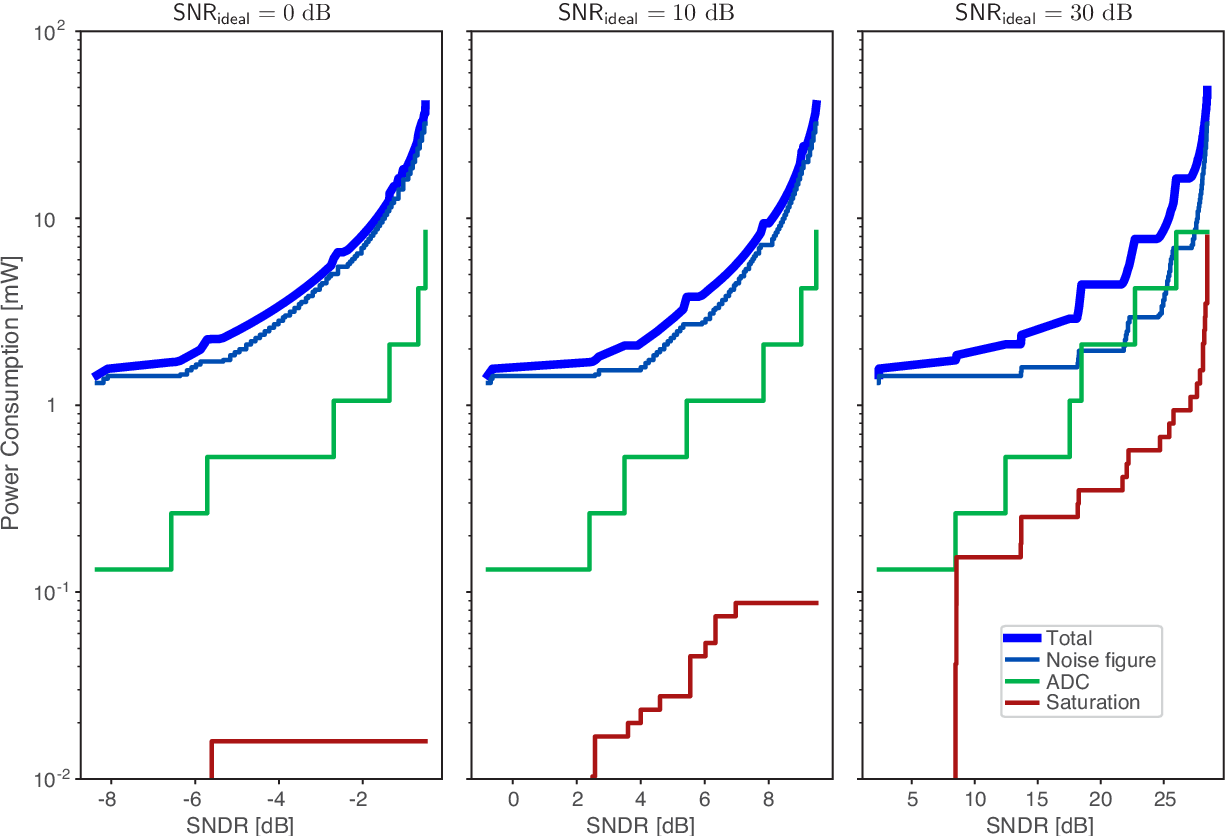}
    \caption{Minimum power consumption as a function of the $\SDNR$, parameterized by $\SNR_{\sf ideal}$, with $\fc=28$ GHz, $B=400$ MHz, and scalar uniform quantization.}  
    \label{fig:pow_comp}
\end{figure*}

Fig.~\ref{fig:pow_comp} breaks the RFE power consumption down into
its three components for the optimum combination of $(F,P_{\sf max},b)$ at every $\SDNR$.
The values increase in discrete steps, reflecting how $(F,P_{\subsf max}, b)$ have been discretized to search for the optimum combinations.
%
Save at the highest $\SNR_{\sf ideal}$,
the saturation power 
is negligible relative to the other components.
Some caution is in order concerning this, as more work is needed to finely model
the behavior of mixers at very low power levels, and also because
the present analysis considers a single link. Additional linear range might have
to be budgeted to handle interference, a point that we return to
later in the article.
%
Nevertheless, 
under the premise that the saturation power can indeed be
neglected, \eqref{Aitana} simplifies to
\be
\cE^{\sf \scriptscriptstyle RFE}_{\rm b} = \frac{1}{\log_2 (1+\SDNR)} \left(
\frac{\gamma_{\sf \scriptscriptstyle NF}}{F-1} \frac{\fc}{B} + \gamma_{\sf \scriptscriptstyle ADC}  2^b \right) ,
\label{Aitana2}
\ee
which is used in the sequel.

\subsection{Noise Figure}

\begin{figure}
	\centering
        \includegraphics[width=\linewidth]{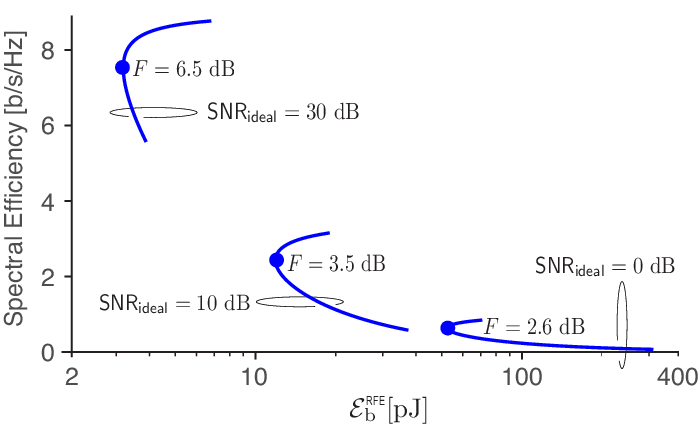}
	\caption{Spectral efficiency vs. RFE energy per bit for
 $\SNR_{\sf ideal} = 0$, $10$, and $30$ dB, with $6$-bit scalar uniform quantization and $\fc/B=70$.  
 For each $\SNR_{\sf ideal}$, the noise figure is varied and the value minimizing $\cE_{\rm b}$ is indicated. }
	\label{Impact_F}
\end{figure}

Also evidenced by Fig.~\ref{fig:pow_comp} is that, when the RFE is optimally configured, the strongest component of the power consumption is virtually always the one associated with the noise figure; physically, this power is invested in improving the quality of the LNA.
Under this premise, it follows from an inspection of (\ref{Aitana2}) that, while relaxing the noise figure worsens the spectral efficiency, it lowers the power consumption even faster, improving the energy per bit.
%
This tradeoff holds up to the point, that depends on the resolution, where the noise-figure power ceases to be dominant; further relaxing the noise figure becomes pointless.
This is illustrated in Fig. \ref{Impact_F}, for $\fc/B=70$ (corresponding for instance to $400$ MHz at $28$ GHz, or to $2$ GHz at $140$ GHz),
with the $\SDNR$ evaluated numerically for $6$-bit scalar uniform quantization. The RFE energy per bit is minimized by $F=2.6$, $3.5$, and $6.5$ dB, respectively for $\SNR_{\sf ideal}=0$, $10$, and $30$ dB. Noise figures above these values yield a simultaneously lower spectral efficiency and energy per bit; below these values, a tradeoff unfolds.


Also interesting in Fig. \ref{Impact_F} is that a higher $\SNR_{\sf ideal}$ improves both the spectral efficiency and the RFE energy per bit. In contrast, the radiated energy per bit worsens with a higher $\SNR_{\sf ideal}$ because of the concavity of the bit rate as a function thereof \cite[Sec. 4.1]{heath2018foundations}. Although beyond the scope of this article, we note that a tension would arise in any holistic optimization of the operating point that involved both transmitter and receiver.

\subsection{ADC Resolution}

Homing in on the ADC, consider first a high $\SNR_{\sf ideal}$.
While increasing the resolution improves the spectral efficiency, the behavior of the RFE energy per bit is more nuanced, and three regimes arise:
\begin{enumerate}
\item For small and moderate resolutions, the power consumption is dominated by the noise figure term, hence the energy per bit actually shrinks as the resolution grows and more bits are pushed through. In this regime, it is pointless not to increase the resolution.
\item At some point, given its exponential dependence on the resolution, the ADC consumption becomes predominant and the energy per bit starts moving north, setting up a tradeoff with the spectral efficiency.
\item Eventually, the spectral efficiency ceases to improve as the performance becomes limited by noise rather than quantization. Past this resolution, energy is squandered for no significant improvement in spectral efficiency.
\end{enumerate}

\begin{figure}
	\centering
    \includegraphics[width=\linewidth]{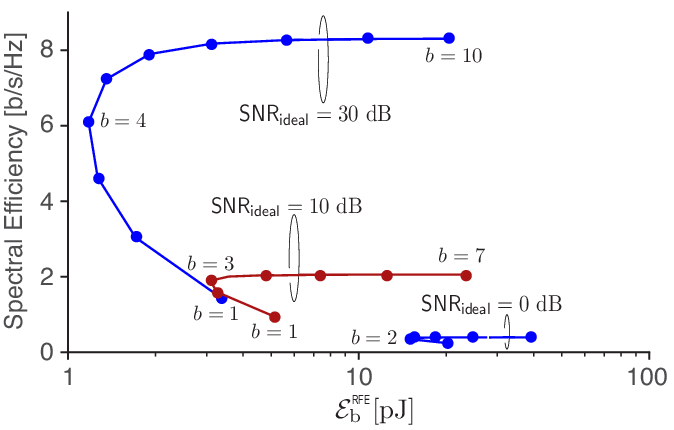}
	\caption{Spectral efficiency vs. RFE energy per bit for
 $\SNR_{\sf ideal}=0$, $10$, and $30$ dB with $F=5$ dB and $\fc/B=70$.
 For each $\SNR_{\sf ideal}$, the resolution of the scalar uniform quantizer
 is varied starting at $b=1$ and the value that minimizes $\cE_{\rm b}$ is indicated.}
	\label{Impact_b}
\end{figure}

Operation in the first and last regimes is ill-advised, and a well-designed system should target the (rather narrow) intermediate one.
This insight would be missed if only the ADC power consumption were accounted for, rather than that of the entire RFE, as the conclusion would then be that the energy per bit is minimized as the resolution is minimized. A more refined analysis reveals that intermediate resolutions are actually preferable.
In particular, the resolution that minimizes the energy per bit is obtainable by solving the transcendental equation that emerges from 
the condition $\frac{\partial \cE_{\rm b} }{ \partial b} = 0$. For $\SNR_{\sf ideal}=30$~dB, $\fc/B=70$, and $F=5$~dB, this returns $b=4$. Shown in Fig. \ref{Impact_b} is how operation below this resolution is indeed unwise while operation above $b=7$ is rather pointless; for $4 \leq b \leq 7$, roughly a four-fold factor in energy per bit can be traded for a roughly $40\%$ increase in spectral efficiency. 


At lower $\SNR_{\sf ideal}$, the same insight holds, only at lower resolutions. For the example in Fig. \ref{Impact_b} translated to $\SNR_{\sf ideal}=10$ dB, a $3$-bit resolution minimizes the energy per bit.
This suggests that an SNR-adaptive resolution would be desirable, and such adaptation is a challenge worth posing to device engineers. Short of that, the resolution must be based on the interval that a system is meant to operate on.

Altogether, the interest in signaling strategies and receiver architectures that are tailored to coarse quantization is well justified, 
not only to prevent excessive power consumption at the receiver, but for the sake of efficiency. In particular, attention should be paid to medium-resolution converters \cite{mezghani2011power}.

\section{Multiantenna Receivers}

Continuing with the case study in the previous section,
let us now graduate to multiantenna receivers, 
progressing from the optimum solution at low SNR (beamforming) to the optimum solution at high SNR (multiple-input multiple-output, or MIMO).

\subsection{Digital Beamforming}

Soaring frequencies and bandwidths sink $\SNR_{\sf ideal}$, because the omnidirectional pathloss scales with $f^2_{\rm c}$ and the noise power scales with $B$. Beamforming is the antidote that can bring things back to the operating range of interest without the need to dramatically shrink the communication distance and/or increase the transmit power.

Digital beamforming at baseband requires an RFE per antenna, followed by a maximal-ratio combiner.
The exact expression for the ensuing $\SDNR$ is somewhat involved \cite{khalili2022quantized}, but, reasonably regarding the distortion at different antennas as uncorrelated, it equals the single-antenna $\SDNR$ multiplied by the number of antennas, $N$. With that, (\ref{curbed}) becomes
\begin{equation} \label{eq:sndr_max_dbf}
    \SDNR \leq  \frac{ N\rho^2(b,\nu)}
    {1-\rho^2(b,\nu)} .
\end{equation}
Likewise, the power consumption would in principle be multiplied by $N$.
As mentioned though, the real appeal of beamforming is that, as the $\SDNR$ is boosted, proportionally lower $\SNR_{\sf ideal}$ can be supported.
Hence, each constituent RFE can operate at a lower resolution, with a lower power consumption.
Weighing the plunging $\SNR_{\sf ideal}$ against the increasing beamforming gain,
and the growing number of RFEs against a lower per-RFE power consumption, 
looms as a most interesting exercise.

\subsection{Analog Beamforming}


To skirt the need for a full RFE per antenna, 
some of the beamforming operations can be conducted in the analog domain, and the resulting architectures are indeed widely used in 5G millimeter-wave
systems. 
The signal received by each antenna is passed through an LNA and a phase shifter. The resulting cophased signals are then added and downconverted, and the ensuing single stream is digitized.
Proceeding as in the single-antenna case, only with $\Ecal_r$ and $\SNR_{\sf ideal}$ multiplied by $N$,
\be \label{eq:sndr_abf}
    \SDNR = \frac{N \, \SNR_{\sf ideal} \, \rho^2(b,\nu)}
    {F + N \, \SNR_{\sf ideal} \, (1-\rho^2(b,\nu))},
\ee
where, rather than \eqref{eq:nudef}, the saturation backoff is now
\begin{equation} \label{eq:nu_abf}
    \nu = \frac{\Ecal_{\sf max}}{N\Ecal_r + N_0} .
\end{equation}
Note that 
\begin{equation} \label{eq:sndr_max_abf}
    \SDNR \leq  \frac{ \rho^2(b,\nu)}
    {1-\rho^2(b,\nu)},
\end{equation}
which, unlike \eqref{eq:sndr_max_dbf}, is unaffected by $N$.
It follows that a higher resolution and saturation level are needed for
the same noiseless SDR as with digital beamforming.
Indeed, as the ADC and mixer are shared by
all $N$ channels, a higher resolution and saturation power are to be expected.

As of the RFE power consumption, it abides by
\begin{equation} \label{eq:Pfe_abf}
    P_{\subsf RFE} = \gamma_{\subsf NF} \frac{N\fc G}{F-1} + \gamma_{\sf max}  P_{\sf max} + \gamma_{\subsf ADC} B \, 2^b,
\end{equation}
where $G$ is an additional LNA gain
required overcome the insertion
loss of the phase shifter. At millimeter-wave frequencies,
a typical value is $G=10$ dB \cite{dutta2019case}.
As only the LNA is replicated $N$ times, only the noise-figure term is affected by the number of antennas, yet this term is now further multiplied by the insertion loss being corrected. And, as mentioned, a higher resolution is needed, meaning that the ADC term is also indirectly enlarged.

\begin{figure}
    \centering
    \includegraphics[width=\linewidth]{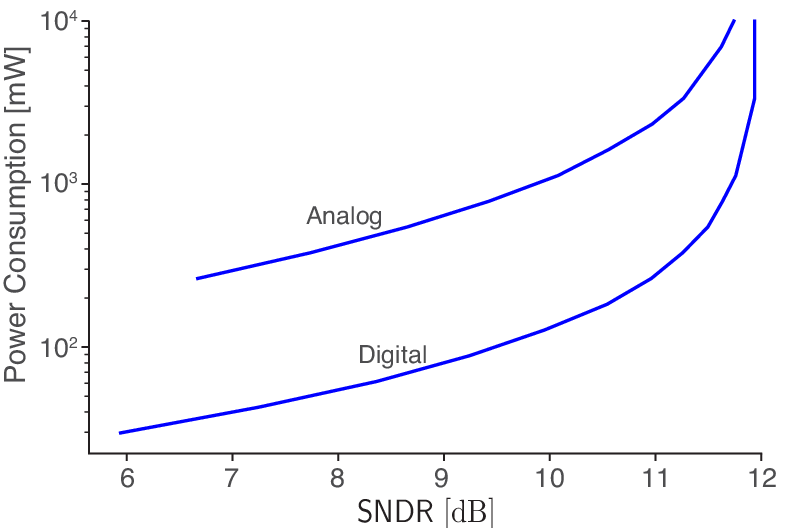}
    \caption{$\SDNR$ after beamforming as a function of the power consumption with $N=16$ receive antennas at $\fc=28$ GHz, $B=400$ MHz, and $\SNR_{\sf ideal}=0$ dB: digital vs analog beamforming.}
    \label{fig:dbf_abf}
\end{figure}

A comparison of the power consumption, digital versus analog,
is presented in Fig.~\ref{fig:dbf_abf}, for $N=16$ antennas and with the configuration of $(F,P_{\sf max},b)$ optimized at every point.
In both cases, having $\SDNR$ approach $N \, \SNR_{\sf ideal}$ (in this case $12$ dB)
entails an escalating power consumption that reaches levels unaffordable for many mobile devices; this exposes the challenge of reaping the full potential of beamforming with $16$ antennas, let alone with the even higher numbers envisioned for 6G.

An important additional conclusion of this analysis is that
the savings in ADC power in the analog structure is more than offset by the increase in LNA power, such that the digital option ends up consuming substantially less power at any given $\SDNR$.
This observation is consistent with studies such 
as \cite{dutta2019case}, yet, as advanced, some caution is in order.
As pointed out earlier, the analysis herein considers that
the saturation power component can be made arbitrarily low,
which would require driving the mixer at very low powers.
While designs such as \cite{skrimponis2020towards} have
used very low power mixers to sacrifice saturation for
power consumption, further work is needed to confirm
that such highly aggressive power savings can be realized.

Digital beamforming is also preferable 
from a performance point of view as it
enables communication in multiple directions
simultaneously; this enables spatial multiplexing as well as dramatically faster
beam search. Also, digital beamforming can mitigate frequency selectivity in the channel, hence its appeal lies beyond the pure spectrum-energy tension.
Our derivations suggest that, looking beyond current commercial designs
that employ analog beamforming to save power,
the benefits of digital beamforming could eventually come with a
power benefit.

\subsection{MIMO}

Finally turning to MIMO, which, like digital beamforming,
necessitates of a full RFE per antenna, it can be abstracted by the combination of a multiplexing gain and a beamforming gain.
There are some caveats concerning the availability of channel-state information available at the transmitter \cite{tulino2004mimo}, and the analysis could branch accordingly, but the multiplexing gain is robust in that respect. 
More nuanced is the dependence on the channel matrix itself, whose singular value spread determines the balance between multiplexing gain and beamforming gain. Having already studied the latter, it is sensible to now focus on situations where the former is maximized; this befits either rich multipath channels at lower frequencies or near-field channels at higher frequencies \cite{do2021terahertz}.
In these situations, at a fixed $\SNR_{\sf ideal}$ both the spectral efficiency and the consumed power
scale with the number of antennas,  
and thus the energy per bit is held constant.
Alternatively, with a diminishing $\SNR_{\sf ideal}$, the spectral efficiency can be held constant while the energy per bit shrinks. 
Both possibilities are decidedly attractive.


\section{1-Bit Communication}

Although, as seen earlier, very low resolutions are not the best choice in terms of energy per bit, and are the worst choice in terms of spectral efficiency, the 1-bit case stands apart for two reasons:
\begin{itemize}
    \item It brings the RFE power consumption down to its minimum, which may be enticing for devices limited, besides battery storage, by sheer power consumption as a proxy for cost, form factor, and heat dissipation.
    \item This extreme approach, in conjunction with 1-bit resolution also at the transmitter, is implementationally attractive: nonlinear power amplifiers can be tolerated at both ends---with the caveat of spectral regrowth and adjacent channel interference---as only the signs of the in-phase and quadrature components matter.
\end{itemize}

Note that dropping the quadrature component could be seen as an even further reduction in resolution, but in actuality it would sacrifice half the bandwidth, hence it is information-theoretically sound to preserve both complex dimensions.

With 1-bit resolutions at the two ends of the link, the single-antenna spectral efficiency is \cite{viterbi1979principles}
\be
C = 2 \left[ 1 - \cH_{\rm b} \! \left( Q ( - \sqrt{\SNR} ) \right) \right]
\label{FCB1}
\ee
where $\cH_{\rm b}(\cdot)$ 
is the binary entropy function and $Q(\cdot)$ is the Gaussian Q-function.
Meanwhile, 
\be
\cE^{\sf \scriptscriptstyle RFE}_{\rm b} = \frac{1}{C}
\left(
\frac{\gamma_{\sf \scriptscriptstyle NF}}{F-1} \frac{\fc}{B} + \gamma_{\sf \scriptscriptstyle ADC}  2^b
\right)
\ee
where, since the spectral efficiency ceases to improve beyond $\SNR \approx 7$ dB, the noise figure can be relaxed. 

The price of the superior energy per bit is, of course, that $C\leq2$ b/s/Hz. This is compounded by the difficulties in engineering a link with such strong nonlinearity early in the processing \cite{singh2009multi}; in that sense, 1-bit communication is a rather unchartered territory, with its own set of challenges.

Beamforming does not remove the $2$-b/s/Hz ceiling,
although it does lessen the $\SNR$ required to attain a certain spectral efficiency. Analog beamforming, in particular, can be straightforwardly subsumed by
using $N \, \SNR$ in lieu of $\SNR$. Digital beamforming, alternatively, requires a generalization of (\ref{FCB1}), depending on the manner in which the signals are combined; from an information-theoretic perspective, the optimum manner is the one that maximizes the mutual information \cite{gao2018beamforming,WSA2021}.

MIMO, if feasible, is the clearest way to higher spectral efficiencies in a 1-bit architecture, and the RFE count is no different from that of digital beamforming: oa full RFE per receive antenna. 
Again, to the extent that the spectral efficiency scales with the number of antennas, the energy per bit does not worsen, making 1-bit MIMO an attractive combination. 

Alternatively to MIMO, there are intriguing information-theoretic results indicating that oversampling can make up for shortages in resolution.
 This is the case if the continuous-time signal is first hard-limited, discriminating it to two levels, and then sampled. The hard-limited signal
is not bandlimited, hence sampling it at the Nyquist rate is suboptimal \cite{gilbert1993increased}. For a properly crafted transmit signal distribution, the $2$-b/s/Hz ceiling can be raised logarithmically in the oversampling factor \cite{shamai1994information,9253719}.
Likewise, if the signal is sampled and subsequently 1-bit quantized, oversampling can be exploited provided the receiver lowpass filter is not set to the signal bandwidth, but expanded by the oversampling factor. Although that increases the noise power, it also ensures that the various samples corresponding to each symbol are contaminated by independent noise and, again for carefully designed signal distributions, a ceiling above $2$-b/s/Hz ceiling can then be attained \cite{krone2012capacity}. Altogether, oversampling can be an interesting alternative to 1-bit communication whenever MIMO is not an option.

No oversampling has been posited throughout this article, and the bandwidth has been identified with the sampling rate. If that is not the case, then 
the sampling rate should be plugged in lieu of the bandwidth in the RFE power consumption formulas.

\section{Conclusion}

As strides are taken towards 6G, energy efficiency is poised to regain its importance alongside spectral efficiency.
To provide information-theoretic cover to this development,
classical formulations 
should be augmented with the power consumed by the circuitry.
At mobile devices in particular, that consumption is to be dominated by the RFE
as we move to higher frequencies, bandwidths, and antenna counts.

Instrumental to any analysis involving the RFE is a model for its power consumption, and this article has set forth such a model. Then, following in the footsteps of other works, the spectral efficiency has been expressed with explicit account of the added noise and distortion introduced by the RFE.
The tradeoff between such spectral efficiency and the energy per bit sheds light
on the effect of the operating point and the FRE configuration. As a case study, receivers with escalating frequencies and bandwidths have been contemplated,
and the gleaned insights include the optimality of intermediate (SNR-dependent) resolutions. The $12$-bit or even $14$-bit resolutions employed by 4G and 5G become decidedly inadequate, and suitably adapting the resolution to the SNR appears as desirable should the hardware be able to accommodate it.
The machinery put forth in the article could be put to use for other situations of interest to 6G. Particularly, multiuser channels with strong near-far conditions
as well as adversarial systems would demand a much higher saturation backoff, bringing into the picture additional distortion and/or power consumption, and possibly altering the contrast of digital and analog beamforming. The corresponding analysis would be a welcome development.

In closing, some final comments are in order:
\begin{itemize}
\item In the face of the non-Gaussian distortion generated by scalar uniform quantizers, and possibly by signal saturation, Gaussian signaling ceases to be optimum. In particular, as far as the quantization distortion is concerned, discrete signal distributions have been shown to be optimum \cite{singh2009limits}.
With optimized discrete constellations, therefore, somewhat better spectral efficiency and energy per bit than the ones considered in this article are possible, yet that need not be the case with generic constellations.
Ongoing research on this matter could augment the framework herein presented.

\item Interference would exhibit a more favorable behavior than thermal noise in that its spectral density would abate as the bandwidth of the interfering transmitters grows; ultimately, the large-bandwidth regime is inherently not limited by interference, but by noise and distortion.

\item Fading and the acquisition of channel-state information are problems on their own right, and they would affect the spectral-energy tradeoff as the bandwidth grows unboundedly---provided the scattering were endlessly rich. However, multipath propagation becomes decidedly sparser as the frequency grows into the realm where very large bandwidths are possible, eroding the inherent channel uncertainty and its impact.
\end{itemize}


A host of research avenues open up as energy efficiency, device power consumption, coarse quantization, distortion, oversampling, and related aspects are injected into information-theoretic formulations. Opportunities to continue probing the boundaries and guiding the evolution of wireless systems, and in particular to keep closing the knowledge gap in terms of how much energy is fundamentally needed to reliably convey one bit of information across a wireless channel.

\iftoggle{arxiv}{}{
\section*{Acknowledgment}

The feedback provided by the guest editorial team and the anonymous reviewers is gratefully
acknowledged.
}



\bibliographystyle{IEEEtran}
\bibliography{jour_short,conf_short,library2,thz_refs,info_theory_refs,sr_refs}

\appendices
\section{Low-Power, High Bit Rate Use Case}
\label{sec:usecases}

An important implication of the analysis herein
is the potential for a high bit rate, short range,
but very low power wireless system.  As an example,
consider the parameters in Table~\ref{tab:linkbudget}, with
a relatively low transmit power of \SI{1}{mW} (similar to Bluetooth low energy). 
The frequency and bandwidth are consistent
with private 5G networks in the CBRS
band
\iftoggle{arxiv}{\cite{wen2021private}}{$\!\!\!$}.
With these parameters, $\SNR_{\sf ideal} = $\,\SI{10.3}{dB}.
Numerically computing $\rho^2(b,\nu)$ and applying
\eqref{eq:sndrsat}, we obtain $\SDNR \approx $\,
\SI{6.0}{dB}; this is about \SI{4.3}{dB} below $\SNR_{\sf ideal}$.
To enable a low power digital implementation,
assume that only 
80\% of capacity at \SI{6}{dB} is attained.
Then, 
$R\approx$\,\SI{160}{Mbps}.

What is remarkable is that the consumption can be small.  Combining Table~\ref{tab:powparam} with \eqref{eq:Pcons},
the total RFE power consumption is below $1$\,\si{mW};
see breakdown in Table~\ref{tab:linkbudget}.
Of course, this
only accounts for the RFE power, to which the digital processing
required for filtering, equalization, and
decoding, must be added.  Nevertheless, this simple calculation
supports the possibility of high bit rate applications
at very low powers.


\begin{table}[t]
 \scriptsize
\begin{center}
  \caption{Theoretical bit rate and power
  consumption for a short range, low power, high rate application.}
  \label{tab:linkbudget}
  \begin{tabularx}{\columnwidth}{|>{\raggedright}>{\hsize=0.4\hsize}X|>{\raggedright}>{\hsize=0.2\hsize}X|
  >{\raggedright}>{\hsize=0.4\hsize}X|}
 \hline
  \rowcolor{NYU_purple!15} 
  Parameter & Value & Remarks 
 \tabularnewline  \hline
 Transmit power & \SI{1}{mW} & 
\tabularnewline  \hline

$\fc$ & \SI{3.5}{GHz} & \multirow{2}{*}{Common in private 5G}
\tabularnewline  \cline{1-2}

$B$ & \SI{200}{MHz} & 
\tabularnewline  \hline

Path loss model & 3GPP InH & Indoor office \iftoggle{arxiv}{\cite{3GPP38901}}{}
\tabularnewline  \hline

Distance & \SI{20}{m} & 
\tabularnewline  \hline

$\SNR_{\rm ideal}$ & \SI{10.3}{dB} & 
\tabularnewline  \hline

$F$ & \SI{4}{dB} & 
\tabularnewline  \hline

$b$ & 4 bits/dim & 
\tabularnewline  \hline

$\nu$ & \SI{30}{dB} &
\tabularnewline \hline 

Spectral efficiency & \multicolumn{2}{l|}{$0.8\log_2(1+0.25 \,\SDNR)$\,[bps/Hz]}
\tabularnewline  \hline

\rowcolor{NYU_purple!10}
\multicolumn{3}{|l|}{Calculated Values for the Bit Rate}
\tabularnewline  \hline

$\SNR_{\sf ideal}$ & \SI{10.3}{dB} &
\tabularnewline  \hline

$\SDNR$ & \SI{6.0}{dB} &
\tabularnewline  \hline

$R$ & \SI{160}{Mbps}  &
\tabularnewline  \hline

\rowcolor{NYU_purple!10}
\multicolumn{3}{|l|}{Calculated Values for the RFE Power}
\tabularnewline  \hline

Saturation power &
\SI{0.003}{mW} & $\gamma_{\rm max}P_{\subsf max}$ 
\tabularnewline  \hline

ADC power  & \SI{0.52}{mW}
&$\gamma_{\subsf ADC}B2^b$
\tabularnewline  \hline

NF power 
& \SI{0.32}{mW} & $\gamma_{\subsf NF}\frac{f_c}{F-1}$ 
\tabularnewline  \hline

$P_{\subsf RFE}$ & \SI{0.84}{mW}
\tabularnewline  \hline

  \end{tabularx}  
  \end{center}
\end{table}

\iftoggle{arxiv}{

\section{RFE Power Consumption Model}
\label{sec:PFEjustification}

This appendix provides details on the figures of merit in Table~\ref{tab:powparam}.

\paragraph*{Noise Figure}

A complete survey 
of published LNA results is furnished in \cite{belostotski_lna_survey}.
For each LNA, one can compute
$\gamma_{\subsf NF} = P(F-1)/f_c$ with the power adjusted
to a typical gain of $G=$\,\SI{15}{dB}.   Shown
in Table~\ref{tab:powparam} is the 25\% percentile
of $\gamma_{\subsf NF}$ for CMOS designs
from 2013 to present and $\fc > $\,\SI{3}{GHz}.

\paragraph*{ADC} A survey of
published ADC performance numbers is maintained in \cite{murmann_adc_survey}.  For each ADC,
one can compute $\gamma_{\subsf ADC} = 2P_{\sf \scriptscriptstyle RFE}/2^b/B$, with the factor of $2$ ensuring that both the in-phase and quadrature ADCs are included,
Table~\ref{tab:powparam} lists the 50\% percentile
of $\gamma_{\subsf ADC}$ for designs
from 2000 with \SI{10}{MHz} $\leq B \leq$ \SI{1}{GHz}.  As mentioned in the body of the article, for sufficiently high bandwidths the ADC power consumption likely grows quadratically with it \cite{murmann2016adc};  however, this behavior need not be considered because the
ADCs can be parallelized.  

\paragraph*{Saturation}
Saturation is most pronounced in the mixer, since this is the component that receives the signal after amplification.  
As there are currently no published surveys on mixers, our sources for mixer data are \cite{wang200951,skrimponis2020towards,guo2014wideband}.
Typically, the saturation level of a mixer 
is quoted by its so-called third order input intercept point (IIP3). For our analysis,
IIP3 needs to be converted to $P_{\sf max}$.  To this end, consider
\eqref{eq:satdef} with 
\begin{equation}
    \phi(u) = u - Du^3 + O(u^4)
\end{equation}
in the vicinity of $u = 0$. For this saturation function,
$    \mathrm{IIP3} = P_{\sf max} / D$ in linear scale.
For $\phi(u) = \tanh(u)$, it can be verified that
$D=2/3$. For each of the designs in 
\cite{wang200951,skrimponis2020towards,guo2014wideband},
we thus computed $P_{\sf max} = 3\mathrm{IIP3}/2$.
Also, since the consumed power should be $P=\gamma_{\sf max}P_{\sf max}$,
the median of $P/P_{\sf max}$ among the designs was taken for
$\gamma_{\sf max}$. 

\section{Derivation of the $\SDNR$ in \eqref{eq:sndrsat}}
\label{sec:sndrsat_deriv}

Since $r=hx$ and $h$ is given,
\begin{equation} \label{eq:rhory}
    \rho_{xy}^2 = \rho_{ry}^2 = \frac{\left|\Exp\left[ r^*y \right] \right|^2}{\Exp|r|^2 \Exp|y|^2}.
\end{equation}
To evaluate $\rho_{ry}^2$, 
let $v = r + z$ be the input to the saturation stage.  Then, $v \sim \Ncgauss(0,\cE_r+N_0)$ and
\begin{equation} \label{eq:yfv}
    y = f(v) = \cQ_b(\alpha S(v)).
\end{equation}
Also, since $(v,r)$ are jointly Gaussian and zero mean, 
we can apply Busgang's theorem to obtain
\begin{align}
    \Exp\left[ r^*y \right] &=
    \Exp\left[ r^*f(v) \right]   \\
    &=
    \frac{\Exp\left[ r^*y\right]}{\Exp|v|^2}\Exp\left[v^*f(v)\right]   \\
    &=
    \frac{\cE_r}{\cE_r + N_0}\Exp\left[v^*f(v)\right] .
    \label{eq:rhory2}
\end{align}
Therefore, 
\begin{align} 
    \rho_{xy}^2 &=
    \frac{\cE_r^2}{(\cE_r + N_0)^2\cE_r\cE_y} \left|\Exp\left[v^*f(v)\right] \right|^2 \\
    &= \frac{\cE_r}{(\cE_r + N_0)^2\cE_y}\left|\Exp\left[v^*f(v)\right] \right|^2   \\
    & = \frac{\cE_r}{(\cE_r + N_0)^2\cE_y}\left|\Exp\left[(r+z)^*y\right] \right|^2 \label{(a)}  \\
    & = \frac{\cE_r}{\cE_r + N_0}
    \rho^2(b,\nu) \label{(b)}    \\
     &= \frac{\SNR_{\sf ideal}}{
     F + \SNR_{\sf ideal}}
    \rho^2(b,\nu)    \label{eq:rhory3}
\end{align}
where 
the definition of $\rho^2(b,\nu)$
in  \eqref{eq:rhobnu} has been applied, along with
\begin{equation}
    \cE_r = \SNR_{\sf ideal} N_0 / F.
\end{equation}
Substituting \eqref{eq:rhory3} into \eqref{SNDR}
we obtain \eqref{eq:sndrsat}.

It remains to prove that $\rho^2(b,\nu)$ is indeed
only a function of $b$ and $\nu$.  Let 
\begin{equation} \label{eq:vtdef}
    \tilde{v} = \frac{1}{\sqrt{\cE_r + N_0}}v,
\end{equation}
such that $\tilde{v} \sim \Ncgauss(0,1)$.
From the definition of the saturation function $S(\cdot)$
in \eqref{eq:satdef},
\begin{align}
    S(v) &= \sqrt{\cE_{\sf max}}\phi \! \left( \frac{|v|}{\sqrt{\cE_{\sf max}}}\right) 
    \frac{v}{|v|}  \\
    & = \sqrt{\cE_{\sf max}}\phi \! \left( \frac{|\tilde{v}|}{\sqrt{\nu}}\right) 
    \frac{\tilde{v}}{|\tilde{v}|}   \\
    & = \sqrt{\cE_{\sf max}}
    S_1(\tilde{v}) \label{eq:SS1}
\end{align}
where we applied \eqref{eq:vtdef} and
\eqref{eq:nudef}, and defined
\begin{equation} \label{eq:S1v}
    S_1(\tilde{v}) = \phi \! \left( \frac{|\tilde{v}|}{\sqrt{\nu}}\right) 
    \frac{\tilde{v}}{|\tilde{v}|}.
\end{equation}
Also, using \eqref{eq:SS1} 
and $v = r+z$, \eqref{eq:alphaq} gives
\begin{align}
    \alpha & = \frac{1}{\Exp[|S(v)|^2]} = \frac{1}{\cE_{\sf max} 
    \Exp[|S_1(\tilde{V})|^2]}.
\end{align}
Hence, for $ \beta = \Exp[|S_1(\tilde{v})|^2]$,
\begin{equation} \label{eq:alphabeta}
    \alpha = \frac{1}{\cE_{\sf max}} \beta .
\end{equation}

Combining \eqref{eq:genrx}, \eqref{eq:SS1}
and \eqref{eq:alphabeta}, the output of the RFE can be written as
\begin{align} \label{eq:yvt}
    y & = \cQ_b(\sqrt{\alpha} S(v) )=   \cQ_b( \sqrt{\beta}S_1(\tilde{v})).
\end{align}
Therefore, in \eqref{eq:rhobnu},
\begin{align}
    \rho^2(b,\nu) & = \frac{\left|\Exp\left[ v^*y \right] \right|^2}{(\cE_r + N_0)\cE_y} \\
   & = \frac{\left|\Exp\left[ \tilde{v}^*y \right] \right|^2}{\cE_y}, \label{eq:rhovt}
\end{align}
where \eqref{eq:vtdef} was also used. 
Now, since $\tilde{v} \sim \Ncgauss(0,1)$, the distribution of $S_1(\tilde{v})$ in \eqref{eq:S1v} is
only a function of $\nu$.  
Thus, the distribution of $y$ in \eqref{eq:yvt}
is only a function of $b$ and $\nu$.
Hence, the expectation in \eqref{eq:rhovt}
can also be expressed as a function of $b$ and $\nu$.

}
{}

\end{document}